# Characterization and correction of time-varying eddy currents for diffusion MRI

Jake J. Valsamis[1,2,*], Paul I. Dubovan[1,2,*], Corey A. Baron[1,2]

[1] Department of Medical Biophysics, Schulich School of Medicine & Dentistry, Western University

[2] Centre for Functional and Metabolic Mapping (CFMM), Robarts Research Institute, London, Ontario, Canada

* JV and PD contributed equally to this work

**Corresponding Author**
Jake J. Valsamis
Email: jvalsami@uwo.ca

**Conflict of Interest**

The authors declare that there is no conflict of interest.

**Funding Information**

Natural Sciences and Engineering Research Council of Canada (NSERC), Grant Number: RGPIN-2018-05448; Canada Research Chairs, Number: 950-231993; Canada First Research Excellence Fund to BrainsCAN; NSERC CGS M program; Ontario Graduate Scholarship program

**Word Count**

4947

# Submitted to Magnetic Resonance in Medicine

# Abstract

**Purpose:** Diffusion MRI (dMRI) suffers from eddy currents induced by strong diffusion gradients, which introduce artefacts that can impair subsequent diffusion metric analysis. Existing retrospective correction techniques that correct for diffusion gradient induced eddy currents do not account for eddy current decay, which is generally effective for traditional Pulsed Gradient Spin Echo (PGSE) diffusion encoding. However, these techniques do not necessarily apply to advanced forms of dMRI that require substantial gradient slewing, such as Oscillating Gradient Spin Echo (OGSE).

**Methods:** An in-house algorithm (TVEDDY), that for the first time retrospectively models eddy current decay, was tested on PGSE and OGSE brain images acquired at 7T. Correction performance was compared to conventional correction methods by evaluating the mean-squared error (MSE) between diffusion weighted images acquired with opposite polarity diffusion gradients. As a ground truth comparison, images were corrected using field dynamics up to third order in space measured using a field monitoring system.

**Results:** Time-varying eddy currents were observed for OGSE, which introduced blurring that was not reduced using the traditional approach but was diminished considerably with TVEDDY and model-based reconstruction. No MSE difference was observed between the conventional approach and TVEDDY for PGSE, but for OGSE TVEDDY resulted in significantly lower MSE than the conventional approach. The field-monitoring-informed model-based reconstruction had the lowest MSE for both PGSE and OGSE.

**Conclusion:** This work establishes that it is possible to estimate time-varying eddy currents from the diffusion data itself, which provides substantial image quality improvements for gradient-intensive dMRI acquisitions like OGSE.

**Keywords**

diffusion MRI, OGSE, eddy currents, validation, field monitoring, brain

# 1 | Introduction

The diagnostic capabilities of diffusion MRI (dMRI) are largely dependent on the gradients used to encode the behavior of water molecules within biological tissue. Advancements in gradient hardware have the potential to broaden the clinical role of dMRI, however eddy currents induced by these strong gradients corrupt data integrity and degrade diagnostic efficacy.[1]

Traditionally, dMRI data is acquired using pulsed gradient spin echo (PGSE), where two gradient lobes with long duration are applied successively.[2] The time between the two gradient lobes and their duration determines the effective diffusion time, which characterizes how long water molecules are allowed to probe their local environment.[3] In PGSE, the effective diffusion time is inherently long, which limits sensitivity to varying microstructural length scales. In contrast, oscillating gradient spin echo (OGSE) has been shown to enable short diffusion times through the use of successive short diffusion weighting periods, with diffusion time scaling inversely with OGSE waveform frequency.[4] Variation of the ADC with OGSE frequency has been observed in the healthy human brain[3,5,6] and OGSE has been used in conjunction with PGSE to provide additional insight into acute ischemic stroke.[7] However, the strong oscillating gradients required for successful OGSE acquisitions necessitate the use of high amplitude gradients with multiple slews. These gradients ultimately induce strong eddy currents that degrade the quality of OGSE data to a greater extent than PGSE.[8,9] Accordingly, a robust technique is needed to correct the induced eddy currents that may be unique to this diffusion weighting method.

There are a variety of precautionary measures taken to compensate for the effects of eddy currents including shielding of magnetic field coils and pre-emphasis of magnetic field gradients. Despite these precautions, eddy current distortions remain a significant burden to dMRI prompting the need for post-processing techniques that can sufficiently correct eddy current corrupted data. A prominent method for correction of eddy currents induced by the diffusion gradients is the "eddy" algorithm in FSL.[10] FSL eddy and other similar packages correct for eddy currents that are assumed to remain constant with time, which leads to simple affine distortions that are relatively straight-forward to detect and correct. While generally effective for PGSE, the applicability of this assumption has not been verified for advanced forms of dMRI, such as OGSE, which have multiple gradient ramps that run the full span from the negative maximum to positive maximum gradient level.

In this work, we utilize dynamic field monitoring (FM) to characterize eddy currents induced by in vivo PGSE and OGSE acquisitions. We then use this characterization to validate a new automatic method that uses the diffusion-weighted data itself to estimate the parameters for an eddy current correction model that, contrary to existing automatic approaches, includes a finite time-constant of eddy current decay. Improved image quality and diffusion dispersion mapping[6] is shown when using this technique compared to FSL eddy.

## 2 | Methods

### 2.1 | MR Acquisition

Two healthy male subjects were scanned on a Siemens MAGNETOM 7T Plus head-only system (80 mT/m gradient strength and 350 T/m/s slew rate). This study was approved by the Institutional Review Board at Western University and informed consent was obtained before scanning. Diffusion MRI data was acquired in a single scan using an in-house sequence consisting of standard PGSE (effective diffusion time 52.8 ms) and cosine modulated trapezoidal OGSE (40Hz). The peak slew rate used was 310 mT/m/s. Signal readout was performed using single-shot EPI with an echo train length of 112, an echo spacing of 0.52 ms, a dwell time of 2.1 μs and a pixel bandwidth of 2125 Hz/Px. The PGSE and OGSE diffusion sequence diagrams are shown in Figure 1. The remaining parameters were b = 750 s/mm$^2$, 4 direction tetrahedral encoding and b = 0 acquisitions with 6 averages each, full Fourier encoding, TE = 124 ms, TR = 11 seconds, field of view = 224 × 224 mm$^2$ , 2 mm isotropic in-plane resolution, 40 slices (2 mm), and scan time 9.5 minutes. The sequence was prepared similarly to the diffusion dispersion sequence optimization parameters that have been previously determined,[6] except here half of the averages were acquired with negative diffusion gradient polarity. Accordingly, for each diffusion-weighted gradient that was applied, another gradient with the same magnitude was applied in the opposite direction to acquire pairs of images that exhibited opposite polarity eddy current distortions. Eddy current compensation was provided by the scanner vendor and calibrated during their normal maintenance.

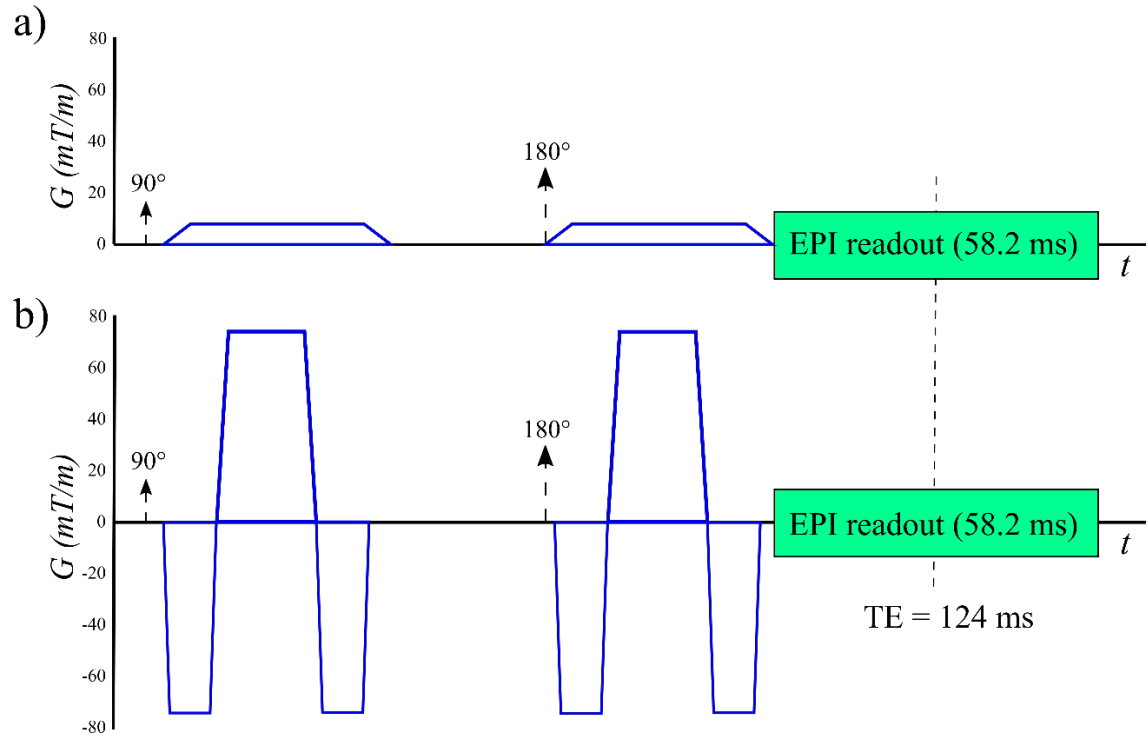

Figure 1

## 2.2 | Eddy Current Characterization and Model-Based Reconstruction

An FM system consisting of 16 $^{19}$F NMR field probes connected to an acquisition system (Skope, Zurich Switzerland) was used to monitor the evolution of field dynamics during the EPI readout period for the same diffusion dispersion protocol performed on the healthy patients. The probes were spherically distributed on a scaffold in verified, fixed positions, and the scaffold was positioned near isocenter in the magnet's frame of reference.[11] Excitation and signal reading of the $^{19}$F probes was handled by inserting a transistor-transistor logic pulse in the MR sequence before the initialization of the readout gradients.

The dynamic spatially varying phase data, which is described with 3rd order spherical harmonics, were computed from the FM raw data using vendor provided MATLAB software. Ignoring the effects of $B_0$ inhomogeneity, the signal measured under the influence of the gradients and diffusion gradient eddy currents is given by:[12]

$$S(t) = \sum_r \rho(r)\exp\left(-i\left[\sum_l \left(k_{l,traj}(t) + k_{l,eddy}(t)\right)b_l(r)\right]\right) \quad (1)$$

where $S(t)$ is the signal acquired at time $t$, $\rho(r)$ is the transverse magnetization at voxel position r, $l$ counts the 16 spherical harmonic terms (0th to 3rd order), $k_{l,traj}$ is the nominal trajectory applied combined with the 1st order eddy currents induced by the EPI gradients, $k_{l,eddy}$ is the phase accrued from diffusion gradient induced eddy currents, and $b_l$ are the spherical harmonic basis functions. The FM system measures $k_{l,traj}(t) + k_{l,eddy}(t)$, and the 1st order terms along the readout and phase encode directions in the non-diffusion weighted scan were subtracted from the corresponding terms in the diffusion weighted scans to estimate $k_{l,eddy}$. Before application of $k_{l,eddy}$ to the data, the spherical harmonic data was interpolated at sampling times corresponding to the halfway point of each phase encode line, where the time per phase encode line was 0.52 ms. Each data point was then replicated to fill its respective readout line. Accordingly, eddy current variation along the readout direction was ignored, which results in substantial time savings during image reconstructions. The measured zeroth order eddy current term was corrected to account for eddy current compensation that is applied to the raw data by the scanner to counteract predicted phase accumulation using off-line simulation of the pulse sequence.

An overview of the reconstruction pipeline is portrayed in Figure 2a, where the eddy current correction step that occurs after N/2 nyquist ghost correction[13] and ramp sampling regridding along the readout direction is either the model-based method described here or TVEDDY (below). For the model-based approach, the k-space data was corrected for diffusion gradient eddy currents using an iterative conjugate gradient solution for $\rho(r)$ using Eq. 1 with $k_{l,traj}$ set to 0 and $k_{l,eddy}$ determined as described above.[14] Note that $k_{l,traj}$ is set to 0 because performing Nyquist ghost correction and regridding of ramp sampling along the frequency encode direction implicitly applied $k_{l,traj}$ already at this point of the reconstruction pipeline. All receiver channels were considered separately, and were combined using a SENSE1 reconstruction[15] after eddy current correction. Receiver combination was followed by PCA denoising applied to the complex data,[16] removal of the phase, export to the NIFTI data standard, Gibbs ringing reduction using the Kellner method,[17] and application of FSL eddy to correct motion between different diffusion weighted acquisitions, which is not corrected by Eq. 1.

### 2.3 | Data Driven Correction: TVEDDY

Eddy currents lead to the k-space trajectory deviating from the intended trajectory, which primarily manifests as distortions and blurring along the phase encode direction due to its slow k-space traversal.[18] Eddy currents are generally spatially and temporally varying in nature. An approximation of the eddy current fields is given by a Taylor expansion of the spatially and temporally varying terms, and a first order approximation typically accounts for the majority of eddy-current induced phase.[19] As such, eddy current fields produced by diffusion gradients can be accurately defined by a spatially invariant term ($B_0$) and linear field gradient terms ($G_x$, $G_y$, and $G_z$), all of which are generally time-varying. Each term is often modelled in time as a sum of exponentially decaying functions with discrete time constants.[19] FSL eddy and similar data-driven approaches assume that there is a single, infinite time constant.

Our approach, which will be referred to as Time-Varying Eddy (TVEDDY) because it considers the time-varying nature of eddy currents, assumes the eddy current fields approximated by the $B_0$ and linear gradient eddy currents are each described by:

$$k_{l,eddy} = A_{\tau,l} exp\{-t/\tau_l\} + A_{inf,l} \tag{2}$$

where $A_{\tau,l}$ and $A_{inf,l}$ are eddy current amplitudes, $\tau_l$ is a finite time-constant, and $l$ = {0, 1, 2, 3} corresponds to $b_l(r)$ = {1, x, y, z}, respectively. It will be assumed henceforth that x, y, and z correspond to the readout, phase encode, and slice-select axes, respectively. For TVEDDY, spherical harmonic terms of 2nd order and higher are ignored. Erroneous bulk distortions (primarily from $A_{inf,l}$) and signal smearing (primarily from $A_{\tau,l}$) that are the result of the eddy currents induced by diffusion gradients will be equal in magnitude but opposite in direction if gradient polarity is reversed.[20] Accordingly, to improve the ability of our approach to use the data to determine a time constant and two separate eddy current amplitudes, TVEDDY currently requires that each diffusion direction is acquired with both positive and negative net diffusion gradient polarity. The optimization procedure aims to determine the eddy current model parameters that minimize the mean squared error (MSE) between images acquired with positive and negative diffusion gradient polarity via multistart gradient descent (MATLAB). Every iteration, the parameters $A_{\tau,l}$, $A_{inf,l}$, and $\tau_l$ are used to estimate $\rho(r)$ from the raw data non-iteratively using a combination of complex phase shifts and nonuniform FFT (Figure 2b). Notably, a non-iterative estimation is possible because only the zeroth and first order terms are considered for TVEDDY, and because the k-space position changes from the diffusion gradient eddy currents are not typically large enough to introduce violations of Nyquist criteria. The eddy current parameters are applied to all slices simultaneously, that is, all slices were assumed to have the same $A_{\tau,l}$, $A_{inf,l}$, and $\tau_l$. The lower threshold of $\tau_l$ was 1.6 ms since smaller time constants decay to zero within a few phase encode lines and the upper threshold of $\tau_l$ was heuristically determined to be 25 ms as larger time constants become close to approximating an infinite time constant (which is already modelled by $A_{inf,l}$).

During a scan, the subject may move between the acquisition of volumes with opposite polarity diffusion gradients. To overcome this limitation, rigid body motion parameters were jointly determined within the optimization to improve the accuracy of the eddy current parameter estimations. PE shifts were omitted from the motion model to avoid redundancy with the $A_{inf,0}$ term, which also creates bulk PE shifts. Finally, our implementation of TVEDDY allows for manually setting $A_{\tau,l}$ to zero, which results in a constant-time correction similar to FSL eddy; for simplicity, this will be referred to as Time-Constant Eddy (TCEDDY). Source code and an example implementation of TVEDDY is available from osf.io/4xtf3/.

The primary purpose of TVEDDY is to regrid trajectory errors that occur due to time varying eddy currents, and it does not compensate for subject motion. To correct for motion, FSL eddy was applied after our custom image reconstruction that contains the TVEDDY algorithm. Also, to mitigate the number of interpolations applied to the data, the final iteration of TVEDDY does not apply the motion parameters. Accordingly, our custom reconstruction uses TVEDDY to correct the in-plane eddy current effects without any interslice interpolation, and all interslice interpolation to correct for motion is applied as a postprocessing step via FSL eddy. Default FSL eddy settings were used with the exception of mporder, an intra-volume motion correction setting, which was set to 4 for all corrections.

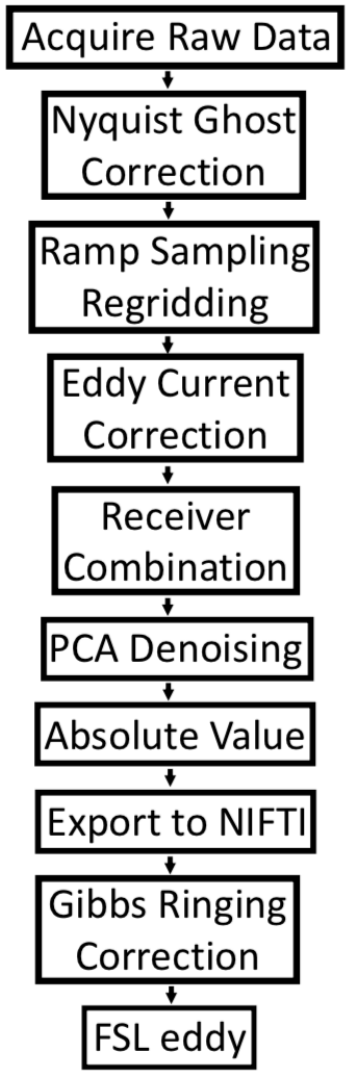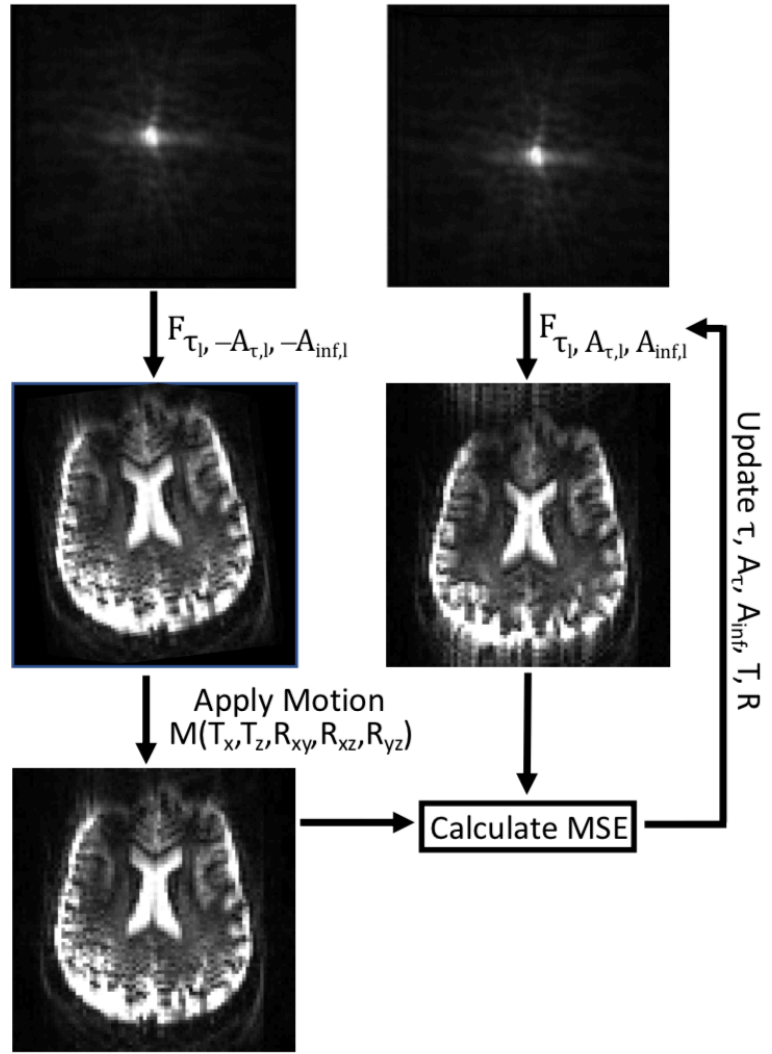

Figure 2

## 2.4 | TVEDDY Validation

Validation 1: Algorithm Convergence. The optimization solved by TVEDDY is generally non-convex, and to assess its ability to find the global minimum, 80 pairs of images were synthetically generated from one of the PGSE diffusion weighted 2D slices acquired *in vivo*. PGSE was chosen for this purpose because time-varying eddy current effects were small for PGSE. Each pair of images was generated assuming eddy currents with opposite polarity and no motion using Eq. 2, where the eddy current parameters $A_{\tau,l}$, $A_{inf,l}$, and $\tau_l$ were randomly generated. After these distortions with known parameters were applied using the TVEDDY forward model (Eq. 2; Figure 2b), noise with a signal to noise ratio of 10 was added to each image. Eddy current correction was performed on a single pair of oppositely distorted slices at a time, and the recovered parameters were correlated against the known parameters to determine if TVEDDY could accurately correct distorted data after the addition of noise.

Validation 2: Model Accuracy. Here, synthetic data was generated from a volume of in vivo PGSE slices. Instead of using Eq. 2, eddy current-distorted raw data was generated on a slice-by-slice basis via Eq. 1 using the zeroth and first order (i.e., $l=\{0,1,2,3\}$) $k_{l,eddy}$ data acquired from an OGSE acquisition using FM, which does not necessarily fit the exponential assumption used in Eq. 2. The volume of simulated images was then processed using TCEDDY and TVEDDY, and eddy model parameters were extracted for comparison of the zeroth and first order eddy current deviations predicted by the two models to the ground truth. Accordingly, this simulation tests the accuracy of the exponential assumption used by TVEDDY. Prior to comparison of TVEDDY to the ground truth eddy current data, global background phase shifts common in both diffusion polarities were removed from the FM data. Given that TCEDDY and TVEDDY only capture inverse eddy current effects, any time varying field terms that are consistent between the two acquisitions will not be accounted for. Accordingly, to separate the common background phase effects from eddy-current induced phase profiles in opposite polarity scans, the profiles were decomposed as follows:

$$k_{l,eddy[pos]} = k_{l[BG]}(t) + k_{l[EC]}(t) \tag{3a}$$
$$k_{l,eddy[neg]} = k_{l[BG]}(t) - k_{l[EC]}(t) \tag{3b}$$

where $k_{l,eddy[pos]}$ and $k_{l,eddy[neg]}$ represent the total FM k-coefficient deviations for positive and negative diffusion gradient directions, $k_{[BG]}$ and $k_{[EC]}$ are the predicted distortions resulting from background and eddy current effects respectively. A zeroth order decomposition example is shown in Figure 3.

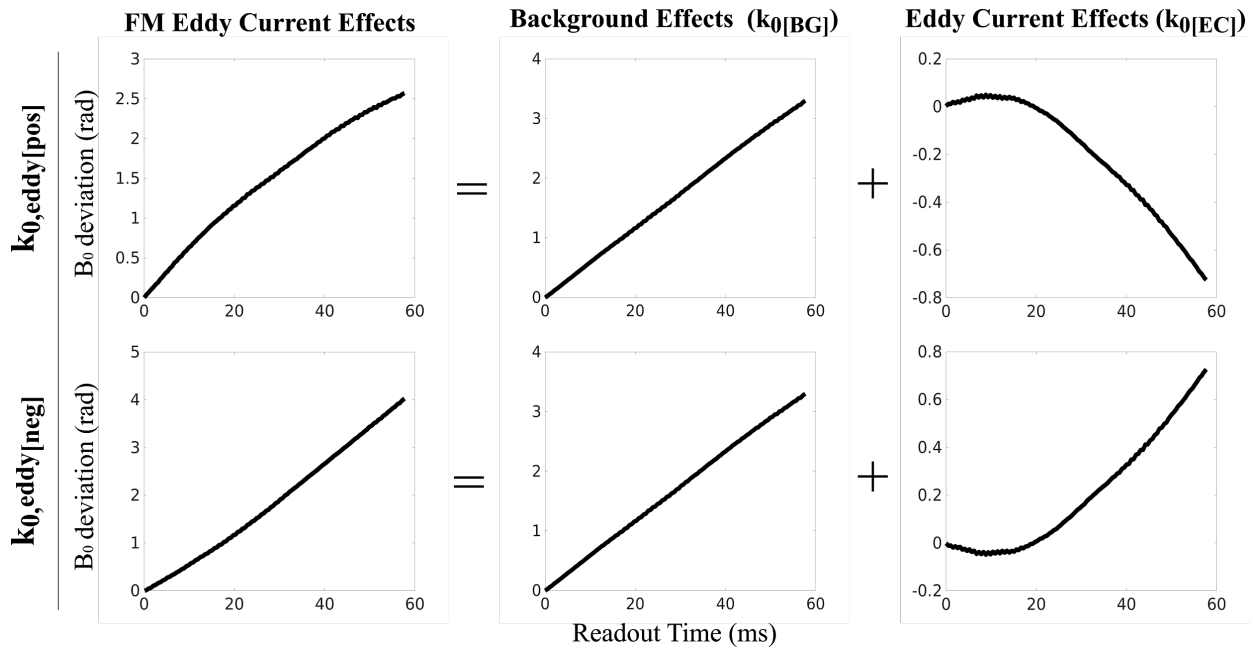

Figure 3

## 2.5 | Technique Comparisons

Reconstructions that used no correction (neither the custom eddy current correction step in Figure 2a nor FSL eddy), FSL eddy only, TCEDDY, TVEDDY, and model-based reconstruction were compared by computing the MSE in pixel intensity between images acquired with opposing diffusion gradient polarities. For the TCEDDY, TVEDDY, and model-based reconstruction cases, FSL eddy was still applied afterward to correct for subject motion; accordingly, any comparisons investigate the incremental gain of combining these eddy current correction approaches with FSL eddy. Applying FSL eddy ensures the motion correction process is consistent for all techniques, reducing the ambiguity in what MSE differences may be attributed to. For each approach and subject, the MSE was measured for all volume pairs with opposing polarity diffusion gradients, normalized by the mean MSE of each subject's PGSE volumes without correction, and averaged over all subjects, diffusion directions, and averages (12 total pairs per subject, each for PGSE and OGSE). A mask made using FSL BET was used to exclude voxels outside the brain. Paired t-tests were used to determine statistical significance between the MSE measurements of different techniques (N=24).

The Diffusion Dispersion (ΔD) is a dMRI parameter that represents the difference in the apparent diffusion coefficient between PGSE and OGSE according to Eq. 4:

$$\Delta D = -\frac{1}{N_\omega} \sum_{i=1}^{N_\omega} \frac{\ln(S_{\omega,i})}{b_{\omega,i}} + \frac{1}{N_{\omega 0}} \sum_{i=1}^{N_{\omega 0}} \frac{\ln(S_{\omega 0,i})}{b_{\omega 0,i}} \quad (4)$$

where $N_\omega$ and $N_{\omega 0}$ are the number of OGSE and PGSE acquisitions, respectively, $S_{\omega,i}$ and $S_{\omega 0,i}$ are the direction-dependent diffusion weighted signals at the frequencies ω > 0 and ω = 0, respectively, and $b_{\omega,i}$ and $b_{\omega 0,i}$ are the direction-dependent b-values at frequencies ω > 0 and ω = 0, respectively.[6] ΔD maps were computed after acquired OGSE and PGSE data were corrected with FSL eddy, TCEDDY, TVEDDY and the model-based reconstruction with FM eddy current data.

## 3 | Results

Generally, PGSE eddy current induced phase increased approximately linearly over the course of the readout, while OGSE exhibited non-linear time dependence (Figure 4).

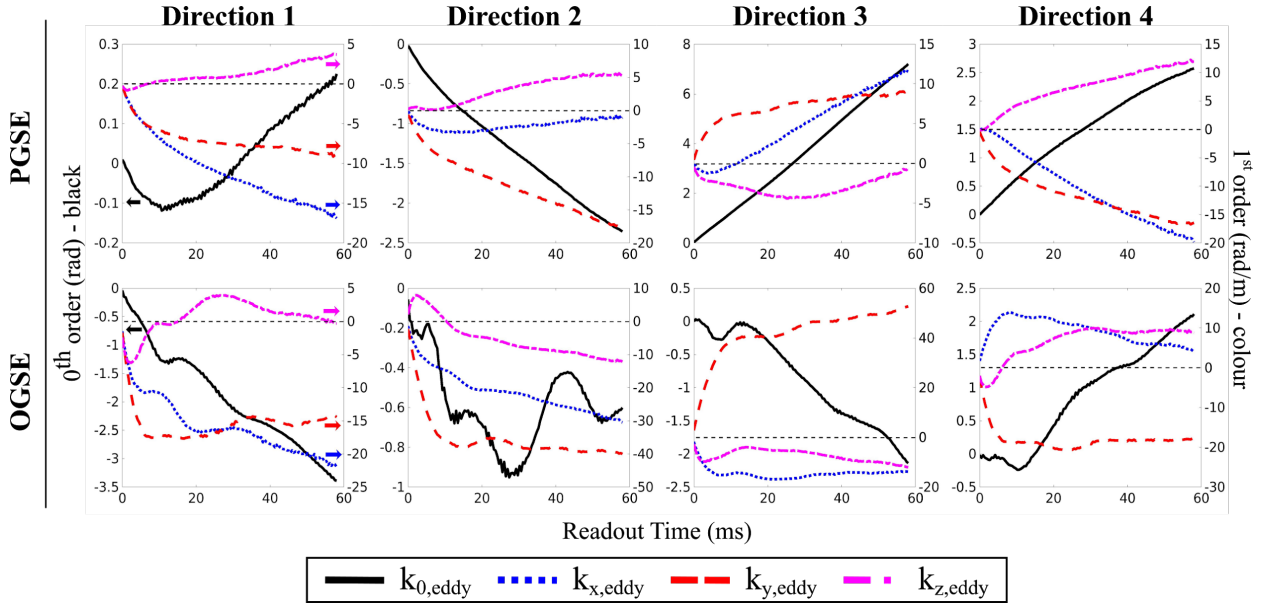

Figure 4

TVEDDY Validation 1: Convergence. Synthetically applied eddy current parameters were accurately recovered using the algorithm's parameter determination capabilities, as indicated in Figure 5 by the good agreement with the expected linear slope of 1 between ground truth and predicted parameters. The slope of the regression line for τ, $A_\tau$, and $A_{inf}$ is 0.689, 0.939, and 0.915, respectively, which indicates a tendency to underestimate the larger values of τ. Bland-Altman analysis revealed little overall bias.

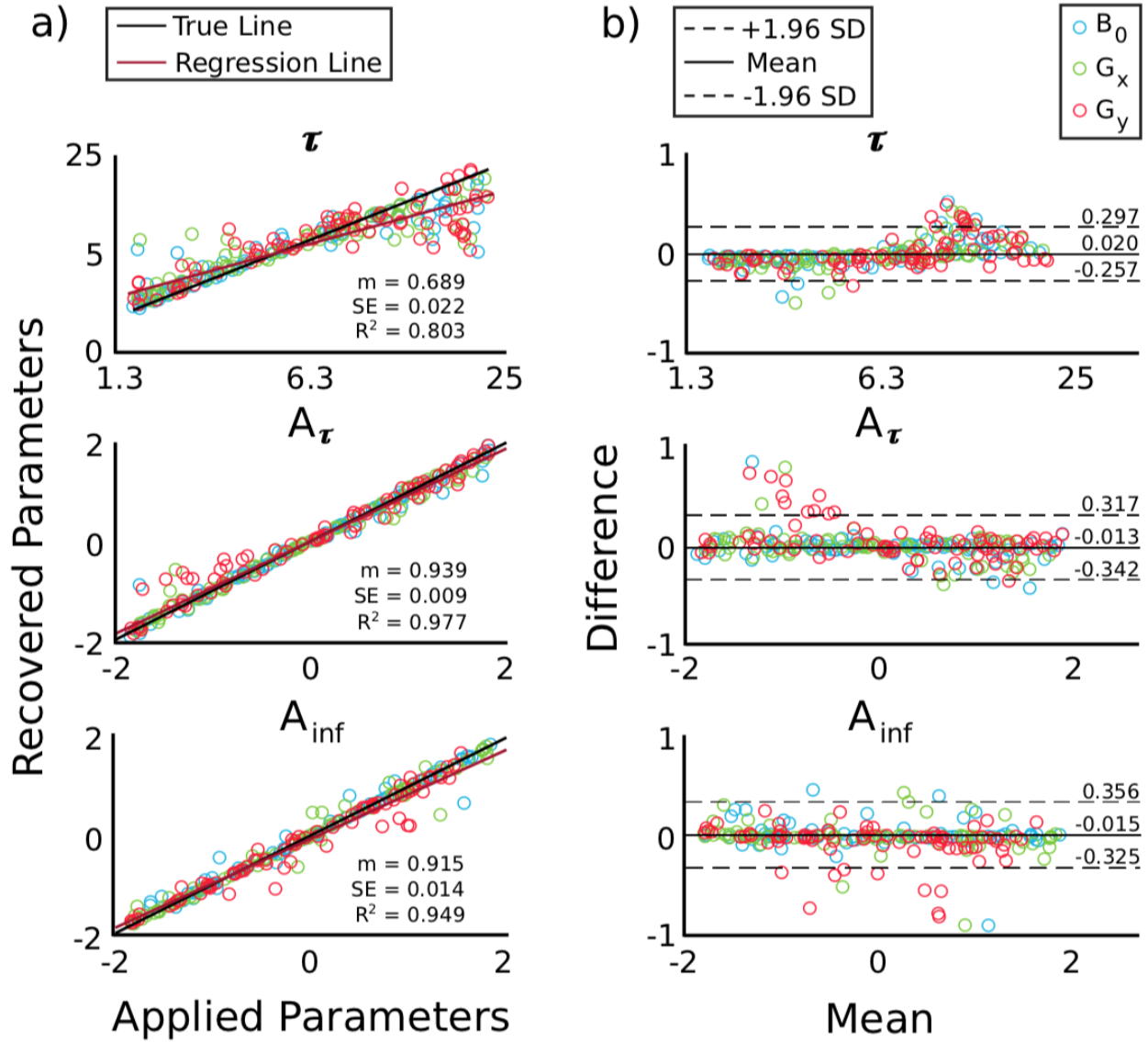

Figure 5

TVEDDY Validation 2: Accuracy. For uniform and relatively linear phase accrual from eddy currents, as typically exhibited by PGSE, the time-constant correction manages to apply phase shifts that reflect the ground-truth phase error well (Figure 6). However, the temporally varying OGSE eddy currents are not well characterized by the time-constant model TCEDDY. The corrections applied using TVEDDY match well with the linear applied corrections for PGSE, while also providing a closer approximation to the more complex eddy currents produced by OGSE compared to TCEDDY.

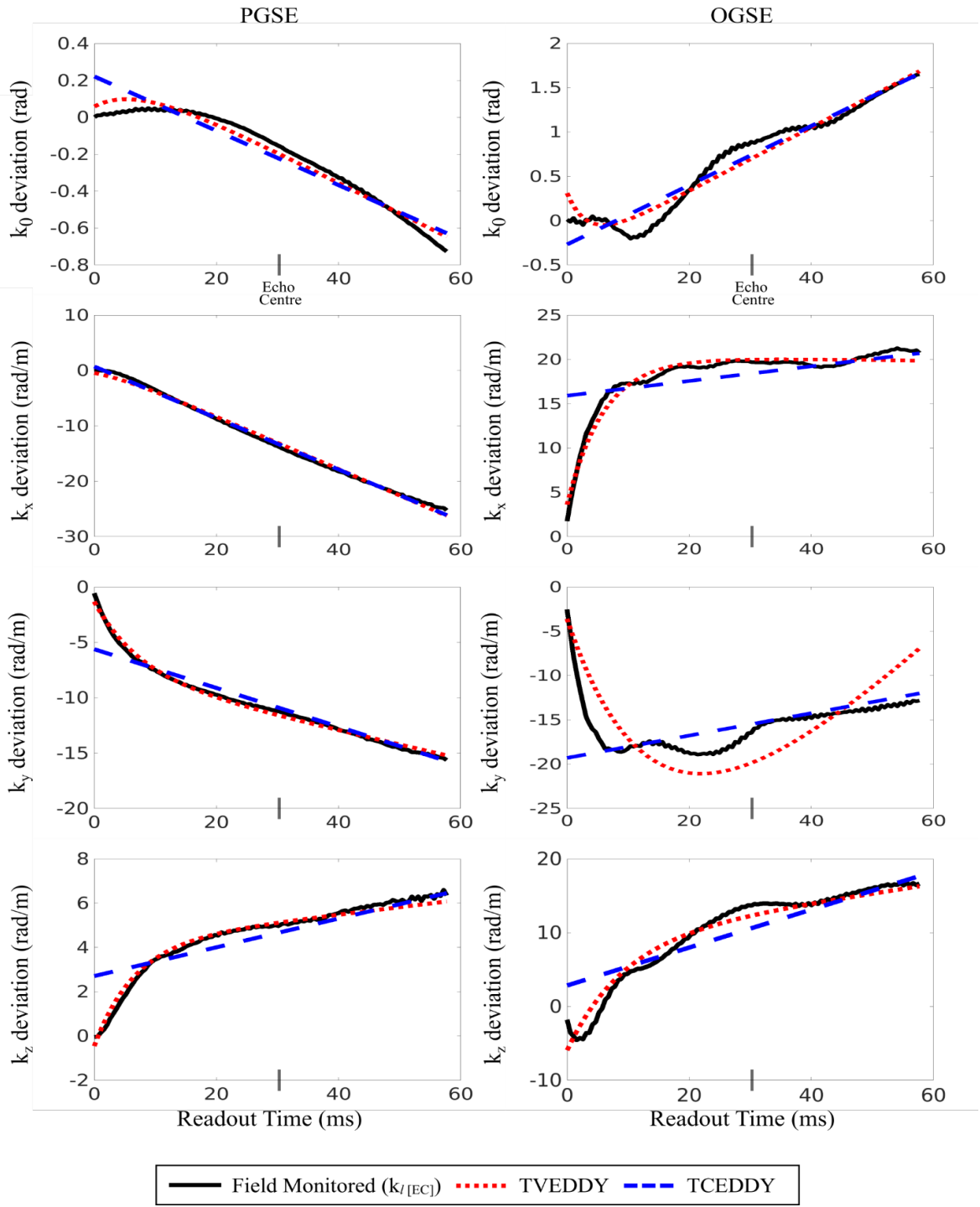

Figure 6

Figure 7a) shows example reconstructed images without correction and the associated difference image after correction with FSL eddy, TVEDDY, FM eddy current modelling. In the original OGSE image, blurring is most noticeable near contrast changes, such as near the ventricles and the cortical ribbon. The difference images show how this blurring persists after correction with FSL eddy but is greatly reduced using the time-varying eddy current model, which is qualitatively comparable to the FM correction.

In Figure 7b), histograms made with each correction technique have the same area, yet the rightward shift in the FSL eddy histogram reveals more voxels with large errors.

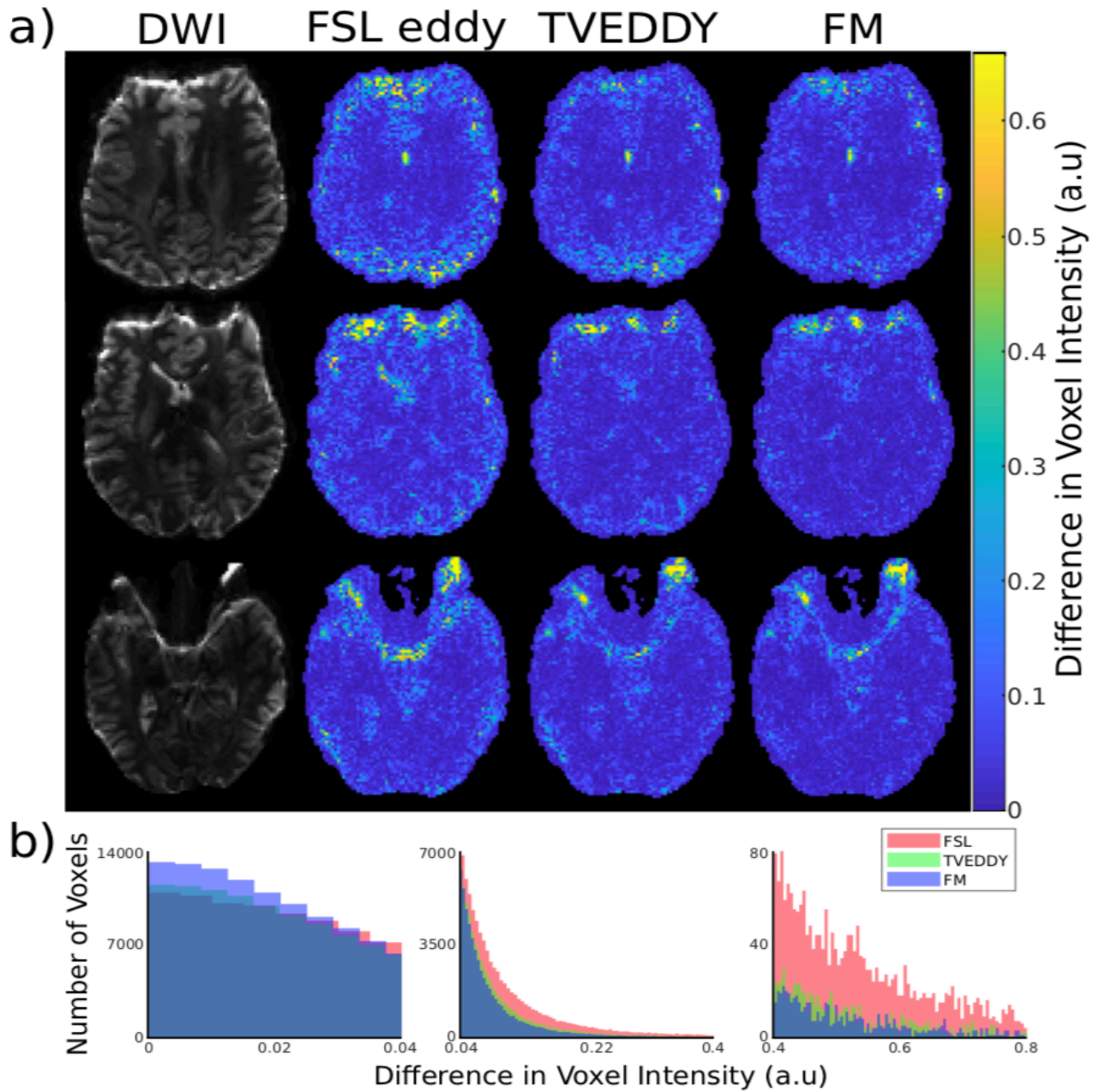

Figure 7

Decreasing MSE values are observed with increasing model complexity (Figure 8). Statistically significant changes in MSE ($P < 0.01$) were observed between each neighboring correction technique in Figure 8, except between TCEDDY and TVEDDY, and TVEDDY and up to first-order FM correction for PGSE volumes.

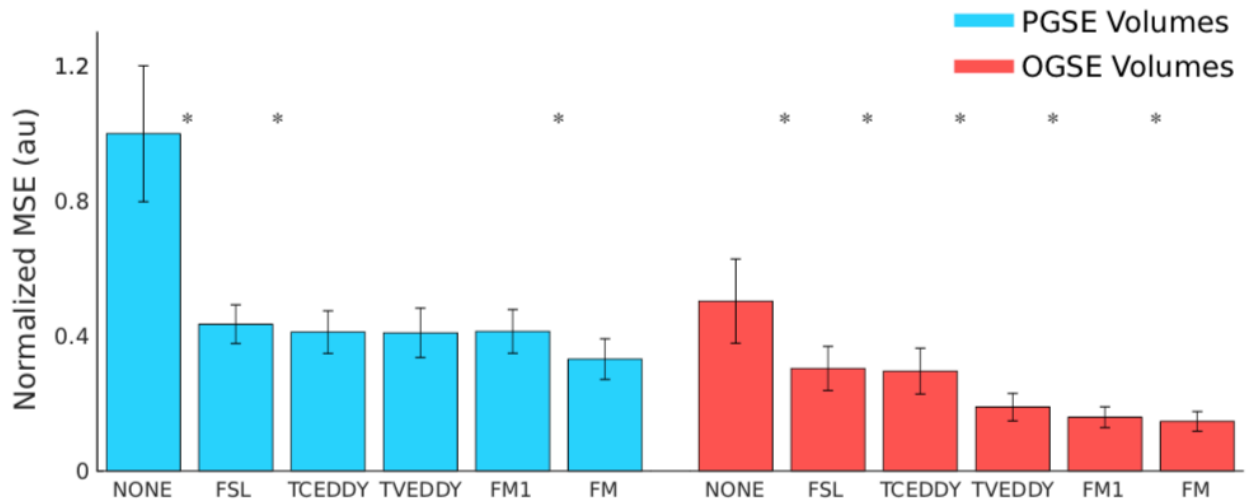

Figure 8

The ΔD maps made using TVEDDY and FM correction show recovered signal voids and improved signal homogeneity of problematic regions in maps made using FSL eddy only, highlighted by yellow circles in both slices of each subject (Figure 9). Notably, the nature of the time-varying eddy currents depends on diffusion gradient direction, which can result in deleterious signal voids when combining the different directions to compute ΔD, and small $\tau$ can create signal pile-up in k-space that results in ringing artefacts that can be observed near the frontal lobe.

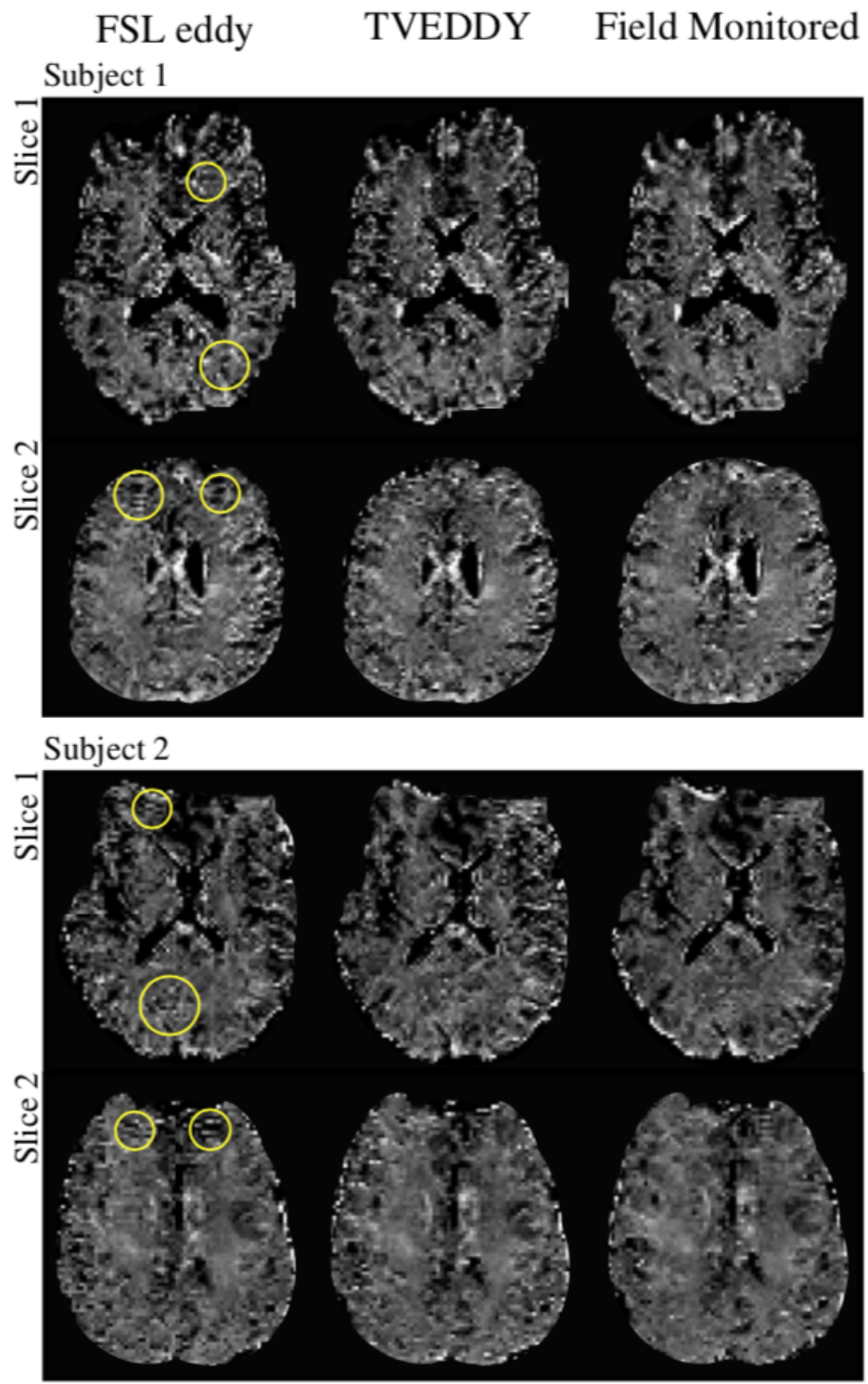

Figure 9

## 4 | Discussion

In this work we have revealed that advanced diffusion weighting approaches may introduce eddy currents that are not compensated well by standard vendor precompensation and introduced a new eddy current correction technique which models diffusion gradient-induced eddy current decay. Current retrospective techniques do not account for this decaying behavior, which results in residual image blurring along the phase encode direction.

The high $R^2$ values and low standard error of the regression line slopes in Figure 5a), and the average difference between applied and recovered parameters being less than or equal to 0.02 in Figure 5b), indicate a consistent relationship between recovered and applied parameters. The $A_\tau$ and $A_{inf}$ regression line slopes close to unity in Figure 5a) suggest that TVEDDY can accurately recover a wide range of eddy current amplitudes. The analysis of recovered parameters reveals an underestimation between recovered and applied $\tau$ that tends to increase with increasing $\tau$. This finding is likely because at large $\tau$ the blurring becomes very subtle and difficult to detect. Our current prototype version of TVEDDY correction of a pair of volumes acquired with opposite polarity diffusion gradients requires up to 1 hour. Thus, a 2D version of TVEDDY that ignores $k_z$ was used to assess convergence which enabled a wide range of eddy current parameters to be tested efficiently. Assessing convergence in 3D may result in slightly different performance, however, TVEDDY's ability to capture eddy current behavior in a volume of data (Figure 6) suggests it can still converge when $k_z$ is included to correct multiple slices.

Looking at the FM zeroth order and first order k-coefficient deviations produced by eddy currents, the PGSE sequences generally exhibited linear behaviour that could be modelled with a single constant term. Conversely, the OGSE scans displayed more complex eddy current evolutions that are not accurately characterized by linear phase accrual. Figure 6 shows how TCEDDY can accurately model the most basic linear phase accruals but performs poorer with any offsets that deviate from linearity. Since TCEDDY is very similar in approach and performance to FSL eddy, it can be inferred that FSL eddy also struggles with correcting distortions from eddy currents with decay constants on the order of the readout duration. Improved blurring correction can be seen with TVEDDY, where including an additional eddy current term with a finite time constant provides additional degrees of freedom to correct for time-varying eddy currents. Overall, this results in correction profiles that more accurately capture the behaviour of applied distortions, which resulted in less blurring/artefacts in reconstructed OGSE images (Figure 7) and ΔD maps (Figure 9), as well as improved similarity between directions acquired with opposite diffusion gradient polarity (Figure 8).

As described in the methods, the displayed FM eddy current-induced profiles in Figure 6 are not results directly monitored by the system, as the original data was decomposed into two time-varying functions, $k_{l[BG]}(t)$ and $k_{l[EC]}(t)$, with $k_{l[EC]}(t)$ being the component that reversed polarity with opposite diffusion gradient polarity (Figure 3). We suspect the background effects arise from temperature increases from the high duty-cycle OGSE diffusion gradients. The temporal zeroth and first order phase accrual originating from the background effects, $k_{l[BG]}(t)$, were approximately linear (Figure 3), primarily affecting alignment of images depending on the diffusion gradient direction, while contributing very little to blurring. Since FSL eddy focuses on correcting linear phase variations, this residual misalignment between non-parallel diffusion directions was likely corrected by FSL in the last step of Figure 2a. Accounting for nonlinear background effects provides almost no improvement to image quality, as seen in correction of PGSE volumes using FM1 relative to TVEDDY, confirming its small contribution.

An unexpected result was the oscillatory eddy current phase accumulation for OGSE (Figure 4). Upon closer investigation, the periodicity of the oscillations is comparable to the period of the OGSE waveforms (~25 ms), suggesting that the gradient oscillations excite oscillatory eddy current modes or that there may be an interaction between residual mechanical vibration of the gradient hardware and eddy current fields. Although TVEDDY provides improved correction by accounting for eddy current decay, it does not capture this oscillating behaviour. Accurately capturing this behaviour would improve correction quality, as seen by the reduced MSE for FM1 recon compared to TVEDDY in OGSE volumes (Figure 8), but the required degrees of freedom for modelling would be high and a robust fitting procedure would likely not be possible. Instead, TVEDDY could be extended to include multiple decaying terms, however, this would increase computation time and may reduce the reliability of correction due to overfitting. While FM can fully account for this oscillating behaviour, it requires dedicated field-monitoring hardware. Alternatively, gradient-impulse response function approaches may be an effective method to account for the eddy currents,[21] but they require detailed knowledge of the diffusion gradient waveforms, which TVEDDY does not. Additionally, it is unclear whether eddy currents produced by OGSE are linear time-invariant, which is a requirement for gradient-impulse response functions to determine an impulse response. Notably, this oscillatory eddy current behaviour may be vendor or hardware specific.

Other strategies for eddy current reduction in OGSE include sinusoid (or near-sinusoid) oscillating gradient ramps and the exclusion of particular encoding frequencies in the OGSE waveform that are responsible for artefacts.[22] However, these techniques come with costs which we could potentially avoid using our method. For example, the exclusion of select frequencies likely requires time consuming calibration and having to choose a lower than optimal frequency would decrease OGSE contrast. The use of non-trapezoidal gradient lobes reduces efficiency in generating desired b-values.[3,22] Notably, these strategies can be used in combination with our method if necessary.

TVEDDY was developed to correct time-varying eddy current distortions that lead to blurring and ringing. This requires pairs of images to be acquired with opposing diffusion gradient polarity which will generally come at no cost to scan time as typical OGSE scans already require many signal averages.[3] FSL eddy was used to correct remaining misalignment between diffusion directions that occur from subject motion and the background eddy current terms (Figure 3). However, it is likely possible to combine the behaviours of TVEDDY and FSL eddy into a single, unified framework. The efficacy of TVEDDY was assessed by determining both quantitatively and qualitatively whether TVEDDY in conjunction with FSL eddy provides any additional benefit to using FSL eddy alone.

Given the opposing polarity of distortions in data acquired with opposite polarity diffusion gradients, volume pairs that are not corrupted by eddy current distortions should have very low MSE (Figure 8). For both PGSE and OGSE volumes, full third-order FM correction resulted in the most improved correction while FSL eddy alone performed the worst out of the tested techniques. In OGSE volumes, TVEDDY outperformed TCEDDY for all volumes acquired. In PGSE volumes, there was no significant difference between correction with TCEDDY, TVEDDY and FM1, which is consistent with the conventional assumption that modelling eddy current decay is not necessary for PGSE, as the linear phase shifts are adequately handled using a static eddy current model. The reduced MSE in OGSE volumes corrected with TVEDDY and FSL eddy versus FSL eddy alone demonstrates the benefit of accounting for eddy current decay when correcting OGSE data. It is also notable that the MSE of PGSE volumes is typically higher than that of OGSE volumes. PGSE gradients tend to induce bulk shifts in the image domain which inflates MSE prior to correction. Additionally, OGSE gradients are intrinsically velocity compensated which ensures there is minimal change in cerebrospinal fluid between OGSE volumes. The substantial movement of cerebrospinal fluid between acquisition of PGSE volumes leads to higher MSE than a corresponding OGSE volume pair once eddy currents have been accounted for. The reduced OGSE MSE can also be attributed to the self-cancellation properties exhibited by OGSE waveforms. Reduced second and third order eddy current effects were observed for OGSE relative to PGSE, consistent with the work presented by Chan et al. for bipolar diffusion gradients.[23] As a result, full third order FM correction provides most of the correction improvement for PGSE volumes while OGSE correction quality is mainly improved using FM1 (see Figure 8).

The ΔD maps illustrate the impact of eddy current correction on a parametric map that can be computed using combinations of OGSE and PGSE scans. Notably, ΔD is vulnerable to eddy current distortions and blurring that present differently between the PGSE and OGSE scans, which is problematic without a correction like TVEDDY given that only the OGSE scans have blurring and ringing due to time-varying eddy currents.

A volume-based correction was used as it provides a good balance between speed and accuracy. The assumption that eddy current parameters do not change across an imaging volume is generally not valid for the first few slices of each TR while the system achieves a steady state, and potentially when the imaging slab size is very large. Otherwise, this assumption provides an accurate definition of eddy currents for an imaging volume. While it is possible to use a slice-based correction approach where each slice is treated separately, this drastically increases the computation time and our preliminary investigations (not shown) suggest that it does not provide substantial improvements in image correction.

The images displayed in this work exhibit distortions from $B_0$ inhomogeneity. Employing parallel imaging would reduce these distortions, but as described by Arbabi et al, the low contrast between PGSE and OGSE volumes necessitates maximizing SNR and also makes ΔD maps particularly sensitive to residual aliasing artefacts that can manifest through parallel imaging.[6] It is likely possible to combine the acquired data with the FM trajectory in a $B_0$ map informed model-based reconstruction to drastically reduce these distortions,[12] but TVEDDY is designed for use in situations where such hardware is not available and it is thus more relevant to validate it in these realistic scanning conditions. Additionally, the implementation of GRAPPA or even partial Fourier may create stronger blurring artefacts due to more data being collected during the most rapidly varying period of the eddy currents. However, the performance of the algorithm would need to be validated for these cases.

FSL topup could also be implemented in a postprocessing pipeline but would require an additional non-diffusion weighted acquisition with reversed phase encoding. This reversed phase encoding scan is distinct from TVEDDY's reversed diffusion gradient requirement. As a result, an additional scan with phase encode gradient reversal (and zero amplitude diffusion gradients) can be performed, and the two correction methods can be applied to the same data to correct both diffusion gradient and image encoding gradient artefacts.

## 5 | Conclusions

This work presented and validated a new computational method for correcting diffusion gradient eddy currents that outperforms FSL eddy for OGSE acquisitions. The capacity to correct distortions induced by advanced dMRI techniques without the substantial cost of an FM system will promote the development and clinical application of advanced dMRI.

**Acknowledgement**

The authors would like to thank the Natural Sciences and Engineering Research Council of Canada (NSERC) [grant number RGPIN-2018-05448], Canada Research Chairs [number 950-231993], Canada First Research Excellence Fund to BrainsCAN, the NSERC CGS M program, and the Ontario Graduate Scholarship program for financial support.

**Data Availability Statement**

The TVEDDY source code along with the necessary files required to correct provided example data is available at https://osf.io/4xtf3/.

**List of Figure Captions**

**Figure 1** The (a) monopolar PGSE and (b) OGSE DW-EPI spin-echo sequences used in the acquisition protocol. The echo time, which is the same for both sequences, is indicated by a dotted line. Diffusion and readout gradient durations are not to scale.

**Figure 2** a) Reconstruction Pipeline. TVEDDY, TCEDDY, or FM eddy current modelling can be selected to perform eddy current correction. b) TVEDDY Algorithm. Eddy current induced artefacts are corrected starting with k-space data that has been acquired with opposite polarity diffusion gradients. The eddy current parameters $A_{\tau,l}$, $A_{inf,l}$, and $\tau_l$, which are applied to the data using phase ramps and a nonuniform FFT (net operator "F"), and translations (T) and rotations (R) are iteratively adjusted until the MSE has converged, where $l$ = {0,1,2,3} corresponds to the zeroth and first order eddy currents (Eq. 2). The data sets shown here were synthetically distorted with large eddy current and motion parameters to illustrate the correction process.

**Figure 3** FM PGSE zeroth order eddy current phase variations decomposed into a common "background" term that does not change when the diffusion gradient polarity is reversed and an "eddy current" term that reverse polarity along with the applied gradients.

**Figure 4** Full EPI readout (58.2 ms) FM zeroth and first order $k_{l,eddy}$ field dynamics along the 4 diffusion gradient directions for PGSE and OGSE acquisitions. Results shown are after the subtraction of FM b = 0 s/mm$^2$ dynamics, but prior to the removal of background effects as per Eq. 3. Zeroth order curves (in black) pertain to the left axis and first order curves (in colour) pertain to the right axis, as indicated by the direction of the arrows. The scale is not consistent between panels.

**Figure 5** a) TVEDDY's ability to recover known eddy current parameters. The linear regression line with slope *m* and standard error *SE* is shown in comparison to the expected true line with a slope of 1.
b) A Bland-Altman plot where the difference between a recovered and applied parameter pair is plotted against the mean of that pair. The solid lines in each plot show the average difference between recovered and applied parameters while the dotted lines represent the interval in which 95% of differences are included. In both panels, $\tau$ is in units of milliseconds and amplitudes are approximately in units of voxels, where $A_{inf}$ = 1 indicates a distortion that causes a shift of one voxel and $A_\tau$ = 1 indicates blurring on the order of one voxel.

**Figure 6** Ground truth FM zeroth and first order eddy current terms (i.e., $k_{l[EC]}$ in Eq. 3) compared to the output of TCEDDY and TVEDDY for one of the four diffusion directions acquired. The entire acquisition window is displayed (58.2ms), and the echo centre has been marked (29.4 ms). Both approaches perform a good correction for PGSE, but only TVEDDY can at least partially correct for the time varying eddy currents observed for OGSE. Similar results were observed for the other three directions.

**Figure 7** a) Qualitative comparison between three eddy current corrupted OGSE slices without eddy current correction, and colourmap difference images between slices acquired with opposite polarity diffusion gradients after correction with FSL eddy, TVEDDY, and FM eddy current modelling. No additional scaling was performed on the difference images relative to the DWI.

b) Three subsets of a histogram showing the frequency of each voxel intensity in a volume of difference images after correction with each technique. Black voxels were excluded from the histogram.

**Figure 8** Quantitative assessment of eddy current correction quality between FSL eddy only (FSL), our time-constant model (TCEDDY), our time-varying model (TVEDDY), and field monitored eddy current correction using up to first-order (FM1) and full third-order (FM) approximation. For both PGSE and OGSE acquisitions, the MSE was measured between all volumes with inverse distortions of both subjects, normalized by the mean PGSE MSE of each subject without correction, and averaged over all directions. Statistically significant changes in MSE between adjacent correction techniques is indicated by * ($P < 0.01$).

**Figure 9** Qualitative comparison between ΔD maps of two slices per subject made with data corrected using FSL eddy only (column 1), data corrected using TVEDDY and FSL eddy (column 2), and data corrected using FM eddy currents and FSL eddy (column 3). Yellow circles highlight problematic regions. The distortions visible in the frontal lobe are due to EPI distortions that stem from $B_0$ inhomogeneity.